\documentclass[journal,twoside,web]{ieeecolor}
\usepackage{tmi}
\usepackage{cite}
\usepackage{amsmath,amssymb,amsfonts}
\usepackage{algorithmic}
\usepackage{graphicx}
\usepackage{textcomp}
\def\BibTeX{{\rm B\kern-.05em{\sc i\kern-.025em b}\kern-.08em
    T\kern-.1667em\lower.7ex\hbox{E}\kern-.125emX}}
\newcommand{\Tone}{T\textsubscript{1}}
\newcommand{\Ttwo}{T\textsubscript{2}}

\newcommand{\ToneTtwo}{T\textsubscript{1}$\rightarrow$T\textsubscript{2}~}

\newcommand{\TtwoTone}{T\textsubscript{2}$\rightarrow$T\textsubscript{1}~}
\newcommand{\TtwoPD}{T\textsubscript{2}$\rightarrow$PD~}

\newcommand{\TtwoFLAIR}{T\textsubscript{2}$\rightarrow$FR~}
\newcommand{\FLAIRTtwo}{FR$\rightarrow$T\textsubscript{2}~}

\newcommand{\ToneTtwon}{T\textsubscript{1}$\rightarrow$T\textsubscript{2}}

\newcommand{\TtwoTonen}{T\textsubscript{2}$\rightarrow$T\textsubscript{1}}

\usepackage{subfig}
\usepackage{tikz}
\usetikzlibrary{matrix,calc}

\definecolor{brightcerulean}{rgb}{0.11, 0.62, 0.74}
\newcommand*{\revhl}{\textcolor{brightcerulean}}

\usepackage[linesnumbered,ruled,vlined]{algorithm2e}
\usepackage{amsmath}
\usepackage{framed,multirow}
\usepackage{fixltx2e}
\usepackage{amssymb}
\usepackage{latexsym}
\usepackage{url}
\usepackage{xcolor}
\usepackage{lipsum}
\usepackage{hyperref}
\SetKwInput{KwInput}{Input}                
\SetKwInput{KwOutput}{Output}              
\hypersetup{
    colorlinks=true,
    linkcolor=blue,
    filecolor=magenta,      
    urlcolor=cyan,
    pdftitle={pFLSynth}
    }
\usepackage{graphics}

\usepackage{gensymb}

\definecolor{newcolor}{rgb}{.8,.349,.1}
\usepackage[font=small,justification=justified,belowskip=2pt,aboveskip=2pt]{caption}
\usepackage[math]{cellspace}
\begin{document}
\title{One Model to Unite Them All: \\ Personalized Federated Learning of Multi-Contrast MRI Synthesis}
\author{Onat Dalmaz, Muhammad U Mirza, Gökberk Elmas, Muzaffer Özbey, Salman UH Dar, Emir Ceyani,\\ Salman Avestimehr, \IEEEmembership{Fellow}, Tolga Çukur$^*$, \IEEEmembership{Senior Member} \vspace{-1.2cm}
 \\
\thanks{\\
This study was supported in part by a TUBITAK BIDEB scholarship awarded to O. Dalmaz, and by TUBA GEBIP 2015 and BAGEP 2017 fellowships awarded to T. \c{C}ukur  (Corresponding author: Tolga \c{C}ukur).}
\thanks{O. Dalmaz, U. Mirza, G. Elmas, M. Ozbey, SUH. Dar and T. Çukur are with the Department of Electrical and Electronics Engineering, and the National Magnetic Resonance Research Center, Bilkent University, Ankara, Turkey (e-mail: \{onat,mumirza,gokberk,muzaffer,salman, cukur\}@ee.bilkent.edu.tr).}
\thanks{E. Ceyani and S. Avestimehr are with the Department of Electrical and Computer Engineering, University of Southern California, Los Angeles, CA 90089 USA (e-mail: \{ceyani@, avestime@\}usc.edu).}
}

\maketitle
\begin{abstract}
Multi-institutional collaborations are key for learning generalizable MRI synthesis models that translate source- onto target-contrast images. To facilitate collaboration, federated learning (FL) adopts decentralized training and mitigates privacy concerns by avoiding sharing of imaging data. However, FL-trained synthesis models can be impaired by the inherent heterogeneity in the data distribution, with domain shifts evident when common or variable translation tasks are prescribed across sites. Here we introduce the first personalized FL method for MRI Synthesis (pFLSynth) to improve reliability against domain shifts. pFLSynth is based on an adversarial model that produces latents specific to individual sites and source-target contrasts, and leverages novel personalization blocks to adaptively tune the statistics and weighting of feature maps across the generator stages given latents. To further promote site specificity, partial model aggregation is employed over downstream layers of the generator while upstream layers are retained locally. As such, pFLSynth enables training of a unified synthesis model that can reliably generalize across multiple sites and translation tasks. Comprehensive experiments on multi-site datasets clearly demonstrate the enhanced performance of pFLSynth against prior federated methods in multi-contrast MRI synthesis.
\vspace{-0.1cm}
\end{abstract}

\begin{IEEEkeywords} federated learning, personalization, MRI, synthesis, translation, heterogeneity, domain shift \vspace{-0.2cm}
\end{IEEEkeywords}

\bstctlcite{IEEEexample:BSTcontrol}

\section{Introduction}
\IEEEPARstart{M}{RI} offers non-invasive assessment of anatomy with rich diagnostic information accumulated over multiple tissue contrasts \cite{bakas2017}. Yet, it suffers from prolonged scan times due to its limited signal-to-noise ratio (SNR) efficiency. Costs associated with multi-contrast protocols can prohibit comprehensive acquisitions or repeat runs of corrupted contrasts during an exam \cite{krupa2015}. Contrast translation is a promising solution wherein missing images in the protocol are synthesized from a subset of acquired images \cite{iglesias2013}. Adoption of deep learning has paved the way for centrally-trained models that offer leaps in synthesis performance \cite{sevetlidis2016,joyce2017,wei2019,pgan}. Unfortunately, training of generalizable models requires diverse datasets that are difficult to curate centrally due to patient privacy risks \cite{Kaissis2020}.

Federated learning (FL) addresses this limitation based on decentralized model training across multiple institutions \cite{WenqiLi2019,Sheller2019,Rieke2020,Roth2020,Li2020,Liu2021}. An FL server sporadically aggregates locally-trained models to compute a shared global model \cite{McMahan2017CommunicationEfficientLO,fl_opt_guide}, and then broadcasts the global model onto each site for further training. Thus, collaborative training is performed on diverse datasets from multiple sites without sharing of imaging data \cite{Li2020FederatedLC}. However, the resultant models can be impaired by the data heterogeneity naturally evident in multi-site datasets \cite{fut_dig_health_fl,FL_medicine}. Previous studies on FL-based medical imaging have introduced prominent approaches to cope with data heterogeneity in segmentation \cite{roth2021,li2021fedbn,Liu2021,fed_dis,fets,fed_semi2021}, classification \cite{Li2020,VAFL,park2021}, and reconstruction \cite{guo2021,feng2021,fedgimp} tasks. However, to our knowledge, no prior study has attempted to address data heterogeneity in federated MRI synthesis. 

Learning-based MRI synthesis recovers unavailable target-contrast images of an anatomy provided as input acquired source-contrast images \cite{vemulapalli2015,jog2017}. Models are trained on datasets comprising samples from the desired source-target configuration \cite{huang2018}. In a restricted setup, a multi-site model can be built for a single task with a common configuration (e.g., \ToneTtwo at all sites where source$\rightarrow$target), resulting in implicit heterogeneity across sites due to variations in sequence parameters or imaging hardware \cite{guo2021,fedgimp}. In a more flexible setup, a multi-site model can be built for multiple tasks with variable configuration across sites (e.g., \ToneTtwo in some, \TtwoPD in others). This will result in explicit heterogeneity due to differences in the identity of source and target contrasts. Both implicit and explicit heterogeneity can induce notable performance losses in synthesis models \cite{pgan}. 

Here, we introduce a novel personalized FL method for MRI Synthesis (pFLSynth) that effectively addresses implicit and explicit heterogeneity in multi-site datasets. pFLSynth employs a unified adversarial model that produces latent variables specific to individual sites and source-target contrasts. To cope with heterogeneity, novel personalization blocks are introduced that adaptively modulate feature maps across the generator given these site- and contrast-specific latents. For further personalization, we propose partial network aggregation on downstream generator layers, while upstream layers are kept local. These design elements enable pFLSynth to reliably generalize across multiple sites and synthesis tasks with a unified model. Comprehensive experiments on multi-site MRI data clearly demonstrate the superior performance of pFLSynth against prior federated models. Code for pFLSynth is available at: {\small \url{https://github.com/icon-lab/pFLSynth}}. 

\vspace{0.2cm}
\revhl{\textbf{Contributions}}
\begin{itemize}
    \item We introduce the first personalized FL method for MRI synthesis to improve performance and flexibility in multi-site collaborations. 
    \item To perform diverse tasks with a unified model, our personalized architecture adaptively modulates feature maps to each site and source-target configuration. 
    \item We propose partial network aggregation on downstream layers to maintain site generality, while upstream layers are kept local to maintain site specificity. 
\end{itemize}

\section{Related Work}
Multi-contrast MRI synthesis has witnessed a recent surge in deep learning methods \cite{bowles2016,chartsias2017,nie2018}. Earlier studies proposed convolutional neural networks (CNN) with pixel-wise loss terms \cite{sevetlidis2016,joyce2017,wei2019}. To improve capture of tissue details, generative adversarial networks (GAN) were also introduced based on adversarial loss terms that indirectly learn the distribution of target images \cite{armanious2018,pgan,yu2019,zhou2020,Shen2,Luping1}. Commonly, maximal performance has been aimed by training a singular model for each source-target configuration \cite{mustgan}. To mitigate computational burden, later studies have instead proposed unified models capable of performing multiple tasks \cite{sharma2019,collagan,li2019,wang2020,resvit}. However, despite demonstrated success in learning-based synthesis, prior studies have predominantly reported centralized models that require cross-site transfer of imaging data \cite{Kaissis2020}. 

To remedy privacy concerns, FL conducts decentralized training via communicating model parameters \cite{Li2020FederatedLC}. Given their potential for collaborative studies, FL methods have been proposed for several imaging tasks including segmentation \cite{Sheller2019,WenqiLi2019,roth2021,Liu2021}, classification \cite{li2021fedbn,Roth2020,VAFL,park2021}, reconstruction \cite{guo2021,fedgimp,feng2021,swfedcyclegan}, and unconditional image generation \cite{federated_synthetic_learning}. Yet, the potential of decentralized procedures in multi-modal medical image translation remains largely unexplored. FL has recently been adopted for multi-contrast MRI synthesis based on cycle-consistent models \cite{fedmedgan,fedmedatl}. Promising results were reported for training singular models built independently for each separate source-target configuration. Yet, to our knowledge, no prior FL study has attempted to address heterogeneity in multi-site datasets during MRI synthesis. 

\begin{figure*}[t!]
\centering
\includegraphics[width=0.99\textwidth]{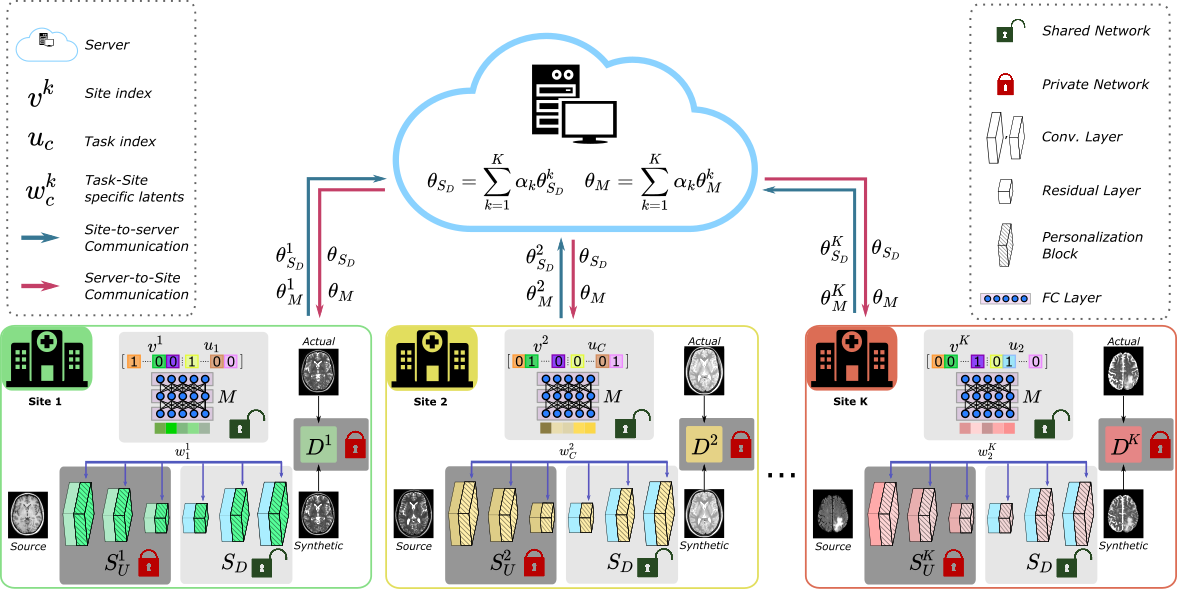}
\caption{pFLSynth is a personalized federated learning model for MRI synthesis. The proposed architecture contains a mapper to produce site- and task-specific latents. Latents modulate feature maps via novel personalization blocks (PB) inserted across the generator. Partial network aggregation is adopted with shared downstream albeit unshared upstream layers to further promote personalization. These design elements enable pFLSynth to offer robust performance under data heterogeneity within and across sites.}
\label{fig:main}
\end{figure*}

Here, we introduce the first FL method for personalized MRI synthesis that addresses heterogeneity existent for both common and variable source-target configurations across sites (Fig. \ref{fig:main}). To do this, pFLSynth introduces novel personalization blocks in the generator of an adversarial model (Fig. \ref{fig:main_generator}). Receiving site- and contrast-specific latent variables, these blocks adaptively tune the statistics of feature maps via adaptive instance normalization \cite{adain}, and tune the attributed importance of feature channels via adaptive channel weighting. Several recent studies have considered adaptive normalization for imaging tasks \cite{ chulye_adain2,adain_chulye_ct_denoising,swfedcyclegan,adainsynthesis1,multipleTEsynthesis}. In \cite{chulye_adain2}, normalization was used for ultrasound imaging to switch among various output types in a centralized model. In \cite{adain_chulye_ct_denoising,swfedcyclegan,adainsynthesis1}, normalization was used to lower model complexity in a cycle-consistent architecture. In \cite{multipleTEsynthesis}, normalization was used to synthesize images at intermediate echo times (TE) via a centralized model. Unlike prior studies, we employ normalization layers to personalize a federated model to each individual site and source-target configuration, and couple them with novel channel weighting layers for improved specificity. 

pFLSynth employs shared generators across sites, albeit the generator is split centrally in its residual bottleneck to enable partial aggregation. A recent study on MRI reconstruction has proposed sharing the encoder in a UNet backbone while keeping the decoder unshared to enhance site specificity in decoded representations \cite{feng2021}. In contrast, here we propose to share the downstream layers of the residual generator while keeping upstream layers local to maintain specificity in encoded representations. Taken together, these unique aspects enable pFLSynth to effectively address heterogeneity in multi-contrast data for reliable MRI synthesis.

\section{Theory}

\subsection{MRI Synthesis with Adversarial Models}
Adversarial models have become pervasive in MRI synthesis due to their sensitivity for high-frequency features \cite{pgan,beers2018,collagan,armanious2019}. For adversarial learning, a generator $G$ synthesizes a target image ($\hat{x}_t=G(x_s)$) given as input a source image ($x_s$), while a discriminator $D$ distinguishes actual ($x_t$) and synthetic ($\hat{x}_t$) target images. Assuming spatially-registered images, a GAN is typically trained to minimize:
\begin{eqnarray}
        \mathcal{L}_{syn}(\mathcal{D},\theta) = \mbox{ } \mathbb{E}_{x_s,x_t}[-(D(x_t) - 1)^{2} - D(G(x_s))^{2} \nonumber \\ + 
        \lambda_{pix} ||x_t - G(x_s)||_{1}],
    \label{eq:generic_gan}
\end{eqnarray}
where $\mathbb{E}$ denotes expectation, $\mathcal{D}$ are training data comprising source-target images, $\theta=\{\theta_G,\theta_D\}$ are model parameters, the first two terms reflect an adversarial loss, the last term reflects a pixel-wise loss with relative weight $\lambda_{pix}$. The traditional approach uses centralized training, where data samples from multiple institutions are aggregated in a single repository \cite{Kaissis2020}. This, however, introduces privacy risks for patients. 

Alternatively, decentralized training can be performed via communication between an FL server hosting a global generator ($G$ with $\theta_G$), and sites keeping local copies ($G^k$ for site $k \in \{1,..,K\}$) \cite{fedmedgan}. Discriminators are typically unshared across sites to minimize risk of information leak \cite{FedGAN}. In each communication round, local copies are initialized with the global model transmitted by the server ($\theta_G^k \gets \theta_G$). Local models are then trained to minimize a local synthesis loss:
\begin{equation}
\label{eq:fed_training}
(\theta^k_G, \theta^k_D) = \underset{\theta^k}{\mathrm{argmin}}\mbox{ } \mathcal{L}^{k}_{syn}(\mathcal{D}^k,\theta^k),
\end{equation}
where $\mathcal{D}^k$ are training data ($x^k_{s_c},x^k_{t_c}$), and $(s_c,t_c)$ denotes the $c$th source-target configuration at site $k$ ($c \in \{1,..,C\}$). After each round, local models are aggregated on the server \cite{McMahan2017CommunicationEfficientLO}:
\begin{equation}
\label{eq:fed_avg}
\theta_G =  \sum_{k=1}^{K} \alpha^k \theta_G^k.
\end{equation}
$\alpha^k$ denote relative site weights typically set to $\frac{n^k}{n}$, where $n$ is the total number of training samples and $n^k$ is the number of training samples at site $k$. The trained global model ($G_{\theta^*}$) is eventually used for local inference:
\begin{equation}
\label{eq:fed_ref}
\hat{x}^k_{t_c} =  G^*(x^{k}_{s_c}).
\end{equation}
Domain shifts due to implicit and explicit data heterogeneity will lead to a compromise model among various sites and synthesis tasks, leading to suboptimal performance.

\subsection{Personalized Federated Learning of MRI Synthesis}
For improved performance in multi-site studies, we propose a personalized FL method for MRI synthesis, pFLSynth (Fig. \ref{fig:main}). pFLSynth leverages novel personalization blocks (PB) composed of Adaptive Instance Normalization (AdaIN) \cite{adain} and novel Adaptive Channel Weighting (AdaCW) layers (Fig. \ref{fig:main_generator}). Partial network aggregation is adopted to further improve site specificity and communication efficiency. 

\subsubsection{Personalized Architecture}
pFLSynth is an adversarial model that receives a source image along with site and source-target configuration information as input. The generator employs a mapper that produces latent variables $w$ for modulating feature maps and a synthesizer for source-to-target translation. 

\par \textbf{Mapper ($M$)}: $M$ is an $L_M$-layer multi-layer perceptron (MLP). Receiving a binary vector for site index ($v^k \in \mathbb{Z}_2^{K}$, $\mathbb{Z}_2=\{0,1\}$) and a binary vector for indices of source and target contrasts ($u_c \in \mathbb{Z}_2^{2C}$), it produces a latent vector:
\begin{equation}
    w^k_c = M(v^k \oplus u_c),
\end{equation}
where $w^k_c\in \mathbb{R}^{J}$, $\oplus$ is concatenation, $J$ is latent dimensionality. Parametrized with $\theta_M$, $M$ produces site- and task-specific latents to drive PBs in the synthesizer.

\par
\begin{figure}[!t]
\centerline{\includegraphics[width=0.47\textwidth]{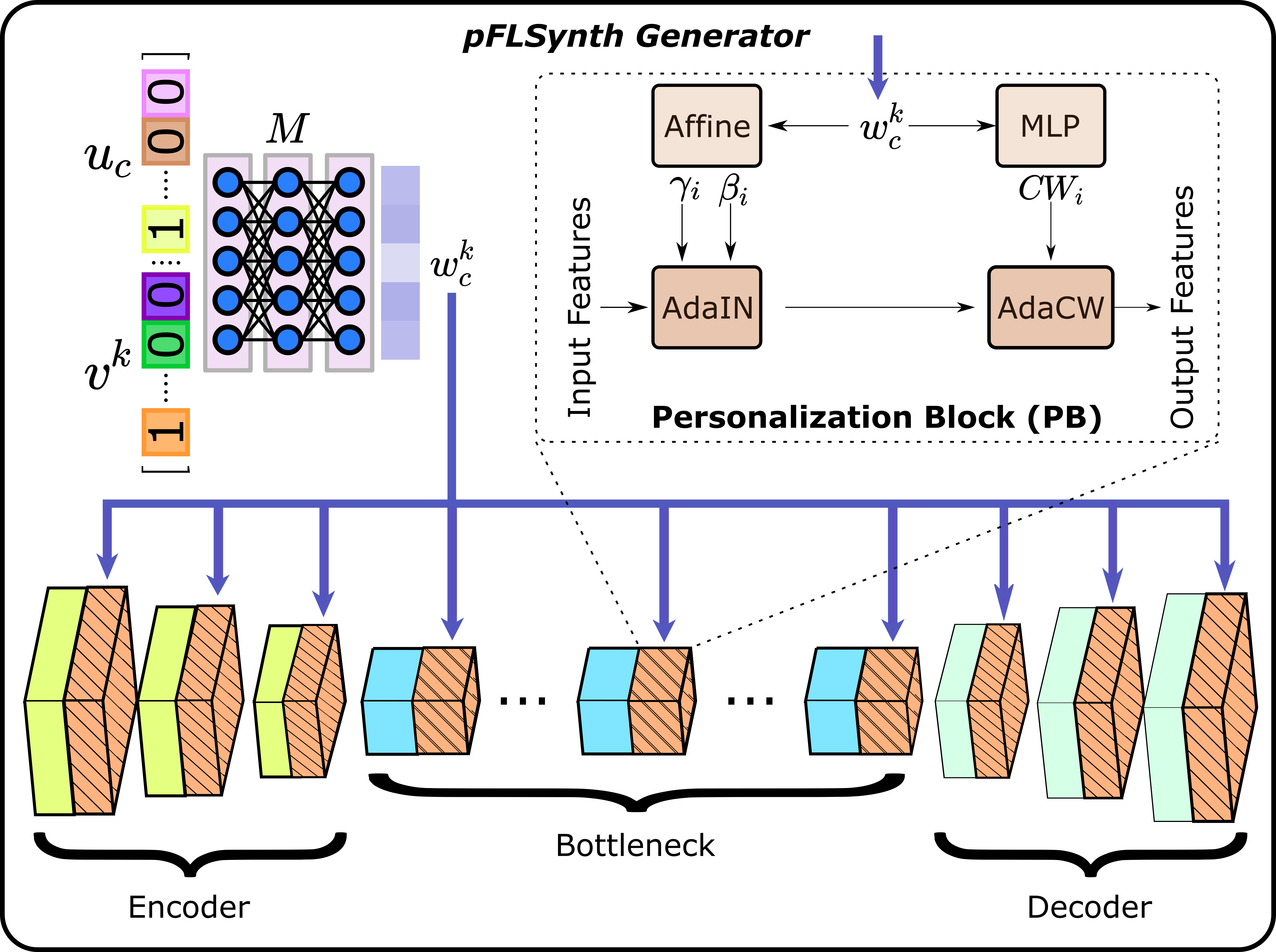}}
 \caption{pFLSynth's generator comprises a fully-connected mapper and a convolutional synthesizer. The mapper produces site- and task-specific latents $w^k_c$, given indices for site identity $v^k$ and source-target configuration $u_c$. The synthesizer computes intermediate feature maps to translate between source and target images. A personalization block (PB) is inserted following each convolutional block that receives $w^k_c$. Each PB adaptively modulates feature maps to alter the statistics via AdaIN and channel weights via AdaCW layers.}
 \label{fig:main_generator}
\end{figure}

 \textbf{Synthesizer (${S}$)}: Receiving as input a source image $x_s$ and latent vector $w^k_c$, $S$ generates a target image $\hat{x}_t$. The backbone is inspired by the ResNet model \cite{pgan,resnet} with a residual bottleneck between an encoder and a decoder (see Fig. \ref{fig:main_generator}). For each site and source-target configuration, synthesizer feature maps elicit divergent statistics across spatial and channel dimensions \cite{cen}. To mitigate this heterogeneity, we introduce novel PBs inserted after each convolutional block in the synthesizer except for the final decoder block. Let $\{ S_{1}, S_{2}, \cdots, S_{L_S} \}$ with parameters $\theta_{S_{1,..,L_S}}$ be the set of synthesizer stages. At the $i$th stage, input feature maps $f_{i-1} \in \mathbb{R}^{F_{i-1},H_{i-1},W_{i-1}}$ ($F_{i-1}$, $H_{i-1}$, $W_{i-1}$ are the number of channels, height and width) are processed via $\mathrm{CB}_{i}$:
\begin{equation}
    g_{i} = \mathrm{CB}_{i} (f_{i-1}) \in \mathbb{R}^{F_{i},H_{i},W_{i}},
\end{equation}
where $\mathrm{CB}_{i}$ denotes the $i$th convolutional block. 

$\mathrm{PB}_{i}$ first transforms $w^k_c$ into scale and bias vectors $\gamma_i,\beta_i \in \mathbb{R}^{F_{i}}$, and then normalizes the statistics of $g_i$:
\begin{eqnarray}
    &\gamma_i (k,c) = Q^{\gamma}_i w^k_c + b^{\gamma}_i; \mbox{ }
    \beta_i (k,c) = Q^{\beta}_i w^k_c + b^{\beta}_i,  \label{eq:mean_aff}\\
&g^{'}_i\mbox{\,=\,}\mathrm{AdaIN}(g_{i},\gamma_i,\beta_i)\mbox{\,=\,}\begin{bmatrix}
      \gamma_i[1] \frac{g_{i}[1] - \mu(g_{i}[1])\mathbf{1}}{\sigma(g_{i}[1])} + \beta_i[1] \mathbf{1} \\ 
      \vdots \\ \gamma_i[F_i] \frac{g_{i}[F_i] - \mu(g_{i}[F_i]) \mathbf{1}}{\sigma(g_{i}[F_i])} + \beta_i[F_i] \mathbf{1} 
\end{bmatrix}
\end{eqnarray}
where $Q^{\gamma,\beta}_i \in \mathbb{R}^{F_{i},J}$ and $b^{\gamma,\beta}_i \in \mathbb{R}^{F_{i}}$ are learnable projections, $\mathbf{1} \in \mathbb{R}^{H_{j},W_{j}}$ is a matrix of ones, $\mu$, $\sigma$ compute the mean and standard deviation of individual channels $g_{i}[j] \in \mathbb{R}^{H_{i},W_{i}}$. 

Next, a novel AdaCW layer weights feature channels according to a site-task relevance vector $CW_{i} \in \mathbb{R}^{F_i}$, produced by an $L_{CW}$-layer MLP given the latents $w^k_c$:
\begin{eqnarray}
    &&CW_{i} (k,c) = \mathrm{MLP} (w^k_c)  \\ 
    &&f_{i}\mbox{\,=\,}\mathrm{AdaCW}(g^{'}_i,CW_{i})\mbox{\,=\,}\begin{bmatrix}
      g^{'}_{i}[1] \odot CW_{i} [1] \mathbf{1}   \\
      \vdots  \\ 
      g_{i}[F_i] \odot CW_{i} [F_i] \mathbf{1} 
\end{bmatrix}
\end{eqnarray}
where $\odot$ denotes Hadamard product, and $f_{i}$ is the output feature map. The overall mapping through the synthesizer is a cascade of projections through CBs and PBs:
\begin{equation}
    \hat{x}_t = \mathrm{CB}_{L_S} \circ
    \mathrm{PB}_{L_S -1} \circ \mathrm{CB}_{L_S - 1} \circ \cdots \circ
    \mathrm{PB}_{1} \circ \mathrm{CB}_{1} (x_s,w^k_c),
\end{equation}
where $\circ$ denotes functional composition, and $S_i := \mathrm{PB}_i \circ \mathrm{CB}_i$ for $i < L_s$ while $S_{L_s} := \mathrm{CB}_{L_s}$ (i.e., the final decoder block).

\par \textbf{Partial Network Aggregation (PNA):}
$S$ is split at stage $L_U$ into two disjoint subsets $S_{U} = \{ S_{1}, \cdots, S_{L_U} \} $ and $S_{D} = \{ S_{L_U + 1},\cdots, S_{L_S} \}$ where $L_U$ is selected from $\{1, \cdots, L_S \} $. The set of upstream stages $S_U$ with parameters $\theta^k_{S_U}, k \in \{1,\cdots, K \}$ are kept locally at each site. In contrast, the set of downstream stages $S_D$ with parameters $\theta_{S_D}$ are shared. Thus, pFLsynth only aggregates $S_D$ to improve site specificity in source-image representations. Here, PNA is only exercised on CBs within $S$. While it can also be adopted for PBs, we observed that using local PBs persistently throughout $S$ performs similarly. Thus, to lower communication costs and potential for information leakage, all PB$_i$ with parameters $\theta^k_{\mathrm{PB}_i}$ where $i \in \{1,\cdots, L_S-1 \}$ are kept locally.   

\par \textbf{Discriminator ($D$):}
A local discriminator $D^k$ with parameters $\theta^k_D$ is trained at site $k$ based on a conditional patch-based architecture \cite{pgan}. Given the source image, $D^k$ estimates the probability that $x$ is an actual target image:
\begin{align}
\label{eq:discrim}
    p_D = D(x,x_s),
\end{align}
where $x$ can be an actual or synthetic target image.

\subsubsection{Federated Training}
Decentralized learning is performed for $P$ communication rounds between the server and individual sites (Alg. \ref{alg:training}). The downstream synthesizer $S_D$ and mapper $M$ are shared across sites, albeit the upstream synthesizer $S_U$ and discriminator $D$ are unshared. In the first round, the server randomly initializes a global generator $G=\{S_D,M\}$ with parameters $\theta_G=\{\theta_{S_D},\theta_{M}\}$. At the start of each round, the server broadcasts the global $S_D$ and $M$ to the sites:
\begin{align}
    \theta^k_{S_D} \gets \theta_{S_D}; \mbox{ } \theta^k_{M} \gets \theta_{M}; k = 1,2,\cdots, K
\end{align}
Local $S_U$ and $D$ are set to their states in the previous round:
\begin{align}
    \theta^k_{S_U} \gets \theta^k_{S_U}; \mbox{ } \theta^k_{D} \gets \theta^k_{D}, 
\end{align}
A local generator is then formed as:
\begin{align}
\label{eq:formgen}
    G^k = \{(S^k_U \sqcup S^k_D),M^k\}.
\end{align}
Next, each local generator is trained for $E$ epochs. Note that pFLSynth consolidates different synthesis tasks within and across sites. Thus, local training data $\mathcal{D}^k$ comprise multiple source-target configurations at site $k$:
\begin{align}
    \mathrm{config}^k = \{(s_1,t_1), \cdots,(s_{C},t_{C}) \},
\end{align}
where a fixed number of configurations $C$ is assumed at each site. Given $\hat{x}^k_{t_c} = G^k(x^k_{s_c},v^k,u_c)$, local models are trained to minimize a compound local synthesis loss across tasks:
\begin{flalign}
    \mathcal{L}_{syn}^k ( \mathcal{D}^k, \theta^k) &= \sum_{c = 1}^{C}  \mathbb{E}_{x^k_{s_c},x^k_{t_c}}[-(D^k(x^k_{t_c},x^k_{s_c}) - 1)^{2} - \nonumber \\ &D^k(\hat{x}^k_{t_c},x^k_{s_c})^{2}  + 
         \lambda_{pix}||x^k_{t_c} - \hat{x}^k_{t_c}||_{1}].
    \label{eq:loss}
\end{flalign}
At the end of a round, each site sends its downstream synthesizer and mapper to the server for aggregation \cite{fedavg}:
\begin{align}
    \theta_{S_D} = \sum_{k = 1}^{K} \frac{n^k}{n} \theta^k_{S_D}; \mbox{ } \theta_{M} = \sum_{k = 1}^{K} \frac{n^k}{n} \theta^k_{M}.    
\end{align}
During inference, each site forms a personalized generator as in Eq. \ref{eq:formgen} to perform contrast translation:
\begin{equation}
\label{eq:pfl_ref}
\hat{x}^k_{t_c} =  G^{k*}(x^{k}_{s_c},v^k,u_c).
\end{equation}
Note that our federated generator performs an adaptive source-to-target mapping at each site and for each synthesis task.

\begin{algorithm}[t]
\small
\DontPrintSemicolon
\KwData{$\{ \mathcal{D}^1, \cdots,\mathcal{D}^K \}$ from $K$ sites}
\KwInput{$P$: number of communication rounds \\
$E$: number of local epochs \\
$G$: global generator with params. $\theta_G = \{ \theta_{S_D}, \theta_M \}$ \\
$G^1, \cdots,G^{K}$: local generators with $\theta_{G^1}, \cdots,\theta_{G^{K}}$ \\
$D^1, \cdots,D^{K}$: local discriminators with $\theta_{D^1}, \cdots,\theta_{D^{K}}$ \\
$Opt()$: optimizer for parameter updates \\
$FedAvg()$: federated averaging \\
}
\KwOutput{$\theta^*_{G^k}$ personalized generators
}
Randomly initialize $\theta_G$ and $\theta_{D^1}, \cdots,\theta_{D^{K}}$ \\
\For{$p = 1$ to $P$}
{
    \For{$k = 1$ to $K$}
    {
    $\theta^k_{S_D} \gets \theta_{S_D}$, $\theta^k_{M} \gets \theta_{M}$ \tcp*{receive global}
    \For{$e = 1$ to $E$}
    {
    Calculate $\nabla_{\theta^k_G} \mathcal{L}_{syn}^k ( \mathcal{D}^k)$ based on Eq. \ref{eq:loss} \\
    $\theta^k_G \gets \theta^k_G - Opt(\nabla_{\theta^k_G} \mathcal{L}_{syn}^k ( \mathcal{D}^k)) $ 
    
    Calculate $\nabla_{\theta^k_D} \mathcal{L}_{syn}^k ( \mathcal{D}^k)$ based on Eq. \ref{eq:loss} \\
    $\theta^k_D \gets \theta^k_D - Opt(\nabla_{\theta^k_D} \mathcal{L}_{syn}^k ( \mathcal{D}^k)) $ 
    }
    }
    $\theta_{S_D,M} \gets FedAvg(\theta^k_{S_D,M})$ \tcp*{aggregate}
    
}
\caption{Training of pFLSynth}
\label{alg:training}
\end{algorithm}
\section{Methods}

\begin{figure*}[t]
\centerline{\includegraphics[width=0.925\textwidth]{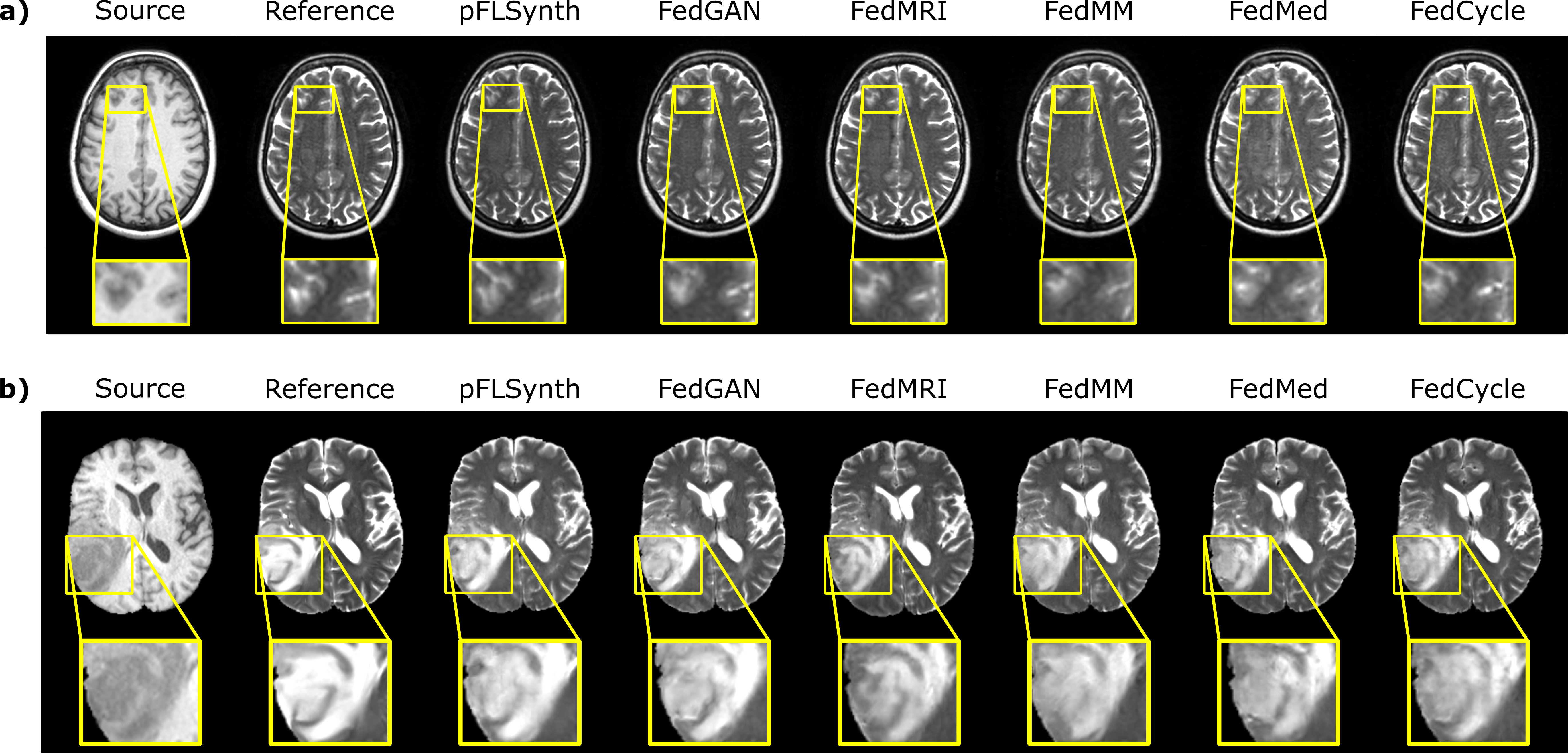}}
\caption{Federated synthesis under a common task configuration within and across sites, \ToneTtwon. Source, reference target, and synthetic target images from competing methods are shown. Representative results are displayed for a) IXI and b) BRATS. Overall, pFLSynth synthesizes images with fewer artifacts and lower noise levels compared to other federated methods.}
\label{fig:exp_1}
\end{figure*}

\subsection{Datasets}
Experiments were conducted on four multi-contrast MRI datasets: IXI\footnote{https://brain-development.org/ixi-dataset/} , BRATS \cite{brats_1}, MIDAS \cite{midas}, and OASIS \cite{oasis}. IXI and MIDAS contain data from healthy subjects, BRATS contains data from glioma patients, and OASIS contains data from subjects with cognitive decline. Each dataset was treated as a separate site in the FL setup. Subjects within each site were split into non-overlapping training, validation, and test sets. In IXI, \Tone-, \Ttwo-, and Proton Density (PD)-weighted images from 53 subjects were analyzed with a (25,10,18) split. In BRATS, \Tone-, \Ttwo-, and Fluid Attenuation Inversion Recovery (FR)-weighted images from 55 subjects were analyzed with a (25,10,20) split. In MIDAS, \Tone- and \Ttwo-weighted images from 66 subjects were analyzed with a (48,5,13) split. In OASIS, \Tone-, \Ttwo-, and FR-weighted images from 48 subjects were analyzed with a (22,9,17) split. Across the four datasets, the training set contained 2780, 2500, 3874, 2780 cross-sections per source-target configuration, respectively. 





\begin{table}[ht]
    \centering
    \footnotesize
    \resizebox{0.40\textwidth}{!}{ 
    \begin{tabular}{ l|c|c|c|c|c|c| }
        \cline{4-7}
        \multicolumn{3}{c|}{} & IXI & BRATS & MIDAS & OASIS \\
        \cline{4-7}
        \multicolumn{3}{c|}{} & \ToneTtwo & \ToneTtwo & \ToneTtwo & \ToneTtwo \\
        \cline{2-7}
        & & P $\Uparrow$ & 28.6$\pm$1.3 & 26.1$\pm$0.9 & 28.1$\pm$0.5 & 25.2$\pm$0.6 \\
        \cline{3-7}
        & Central & S $\Uparrow$ & 94.3$\pm$1.3 & 93.0$\pm$1.1 & 91.9$\pm$0.9 & 83.7$\pm$2.4 \\
        \cline{3-7}
        & & F $\Downarrow$ & 7.4 & 24.8 & 9.7 & 18.1 \\
        \cline{1-7}
        \multicolumn{1}{|l|}{\multirow{18}{*}{\rotatebox[origin=c]{90}{\textit{Federated models}}}} && P $\Uparrow$ & \textbf{28.6$\pm$1.3} & \textbf{26.3$\pm$1.0} & \textbf{28.2$\pm$0.5} & \textbf{25.1$\pm$0.6} \\
        \cline{3-7}
        \multicolumn{1}{|l|}{}& pFLSynth & S $\Uparrow$ & \textbf{94.4$\pm$1.4} & \textbf{93.3$\pm$1.1} & \textbf{92.1$\pm$0.8} & \textbf{84.4$\pm$2.0} \\
        \cline{3-7}
        \multicolumn{1}{|l|}{}& & F $\Downarrow$ & \textbf{7.7} & \textbf{23.9} & \textbf{9.2} & \textbf{30.2} \\
        \cline{2-7}
        \multicolumn{1}{|l|}{}& & P $\Uparrow$ & 27.6$\pm$1.1 & 26.0$\pm$0.9 & 27.9$\pm$0.6 & 24.6$\pm$0.7 \\
        \cline{3-7}
        \multicolumn{1}{|l|}{}& FedGAN & S $\Uparrow$ & 93.2$\pm$1.3 & 92.9$\pm$1.1 & 91.6$\pm$0.8 & 83.4$\pm$2.6 \\
        \cline{3-7}
        \multicolumn{1}{|l|}{}& & F $\Downarrow$ & 11.5 & 24.2 & 9.9 & 40.2 \\
        \cline{2-7}
        \multicolumn{1}{|l|}{}& & P $\Uparrow$ & 27.9$\pm$1.2 & 25.7$\pm$0.8 & 27.7$\pm$0.5 & 23.8$\pm$0.8 \\
        \cline{3-7}
        \multicolumn{1}{|l|}{}& FedMRI & S $\Uparrow$ & 93.7$\pm$1.3 & 92.9$\pm$1.1 & 91.2$\pm$1.0 & 82.7$\pm$2.1 \\
        \cline{3-7}
        \multicolumn{1}{|l|}{}& & F $\Downarrow$ & 9.6 & 32.5 & 11.3 & 43.0 \\
        \cline{2-7}
        \multicolumn{1}{|l|}{}& & P $\Uparrow$ & 26.9$\pm$1.1 & 25.6$\pm$0.8 & 27.9$\pm$0.6 & 24.3$\pm$0.5 \\
        \cline{3-7}
        \multicolumn{1}{|l|}{}& FedMM & S $\Uparrow$ & 91.7$\pm$1.9 & 92.6$\pm$1.1 & 91.5$\pm$0.9 & 80.0$\pm$2.4 \\
        \cline{3-7}
        \multicolumn{1}{|l|}{}& & F $\Downarrow$ & 18.1 & 28.3 & 9.5 & 42.3 \\
        \cline{2-7}
        \multicolumn{1}{|l|}{}& & P $\Uparrow$ & 27.1$\pm$1.1 & 25.6$\pm$0.7 & 27.8$\pm$0.6 & 24.3$\pm$0.5 \\
        \cline{3-7}
        \multicolumn{1}{|l|}{}& FedMed & S $\Uparrow$ & 92.5$\pm$1.4 & 92.5$\pm$1.1 & 91.3$\pm$0.8 & 81.0$\pm$1.9 \\
        \cline{3-7}
        \multicolumn{1}{|l|}{}& & F $\Downarrow$ & 13.0 & 26.6 & 9.4 & 47.0 \\
        \cline{2-7}
        \multicolumn{1}{|l|}{}& & P $\Uparrow$ & 27.2$\pm$1.1 & 25.5$\pm$0.8 & 27.8$\pm$0.6 & 24.2$\pm$0.5 \\
        \cline{3-7}
        \multicolumn{1}{|l|}{}& FedCycle & S $\Uparrow$ & 92.8$\pm$1.5 & 92.4$\pm$1.1 & 91.5$\pm$0.9 & 80.8$\pm$2.2 \\
        \cline{3-7}
        \multicolumn{1}{|l|}{}& & F $\Downarrow$ & 13.4 & 25.5 & 9.8 & 45.8 \\
        \hline
    \end{tabular}
    }
    \caption{Performance in a common task within and across sites (i.e., \ToneTtwon). A centrally trained benchmark is also reported. PSNR (P, dB) and SSIM (S, \%) are listed as mean$\pm$std across test subjects, whereas FID (F) summarizes the entire test set. Boldface indicates the top-performing federated model for each site, task, and metric.}
    \label{tab:unidirectional}
\end{table}

\begin{figure*}[t]
\centerline{\includegraphics[width=0.925\textwidth]{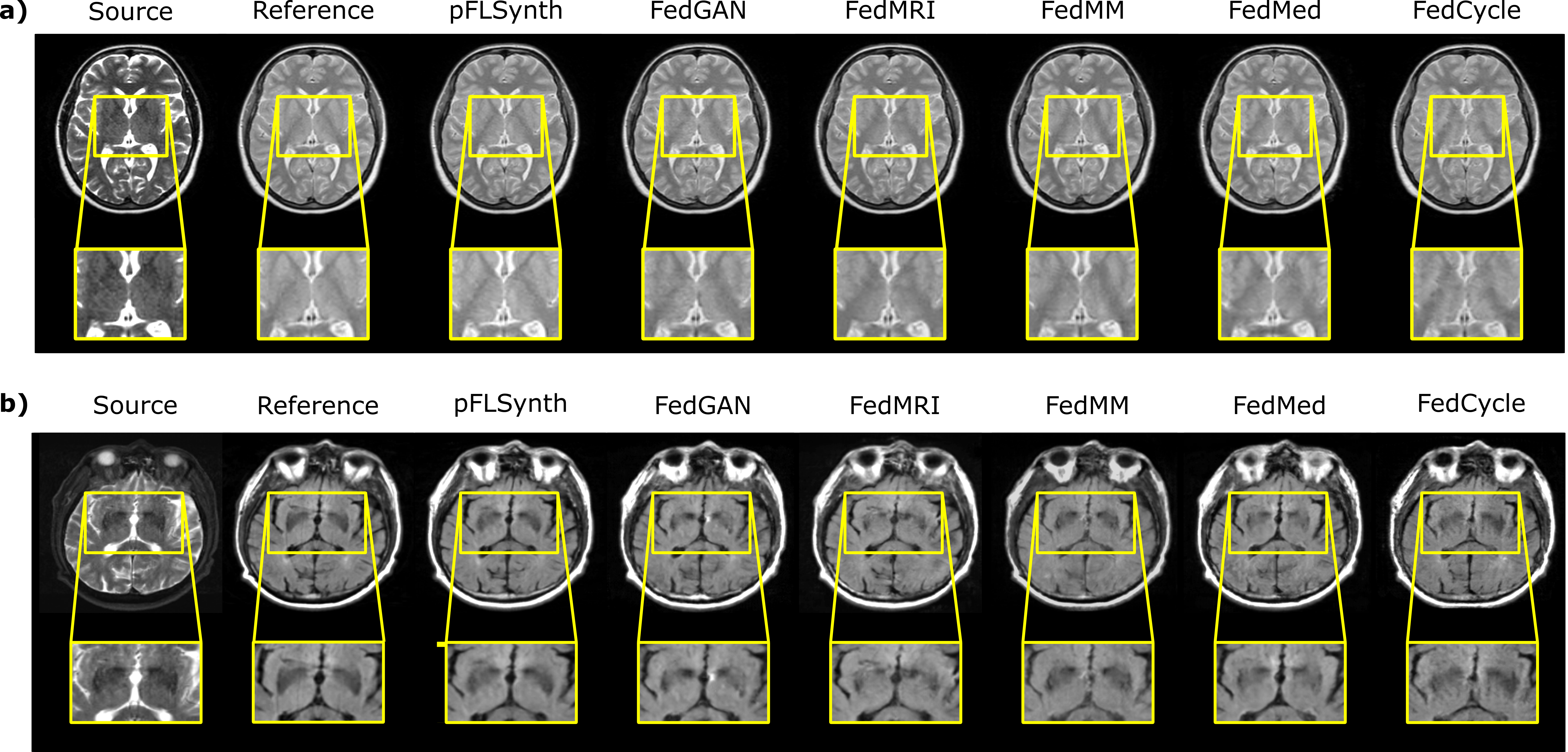}}
\caption{Federated synthesis under a variable task configuration within and across sites (\TtwoTone~and \TtwoPD in IXI, \TtwoTone and \FLAIRTtwo in BRATS, \ToneTtwo and \TtwoTone in MIDAS, \TtwoTone and \TtwoFLAIR in OASIS). Source, reference target, and synthetic target images from competing methods are shown. Representative results are displayed for a) \TtwoPD in IXI and b) \TtwoFLAIR in OASIS. Overall, pFLSynth synthesizes images with fewer artifacts and lower noise levels compared to other federated methods.}
\label{fig:exp_3}
\end{figure*}

\subsection{Competing Methods}
We demonstrated pFLSynth against a centralized benchmark \cite{pgan} and federated baselines \cite{FedGAN,fedmri,fedmedgan,swfedcyclegan,mmgan}. For each method, hyperparameter selection was performed via identical cross-validation procedures. All models shared generators across sites but used a separate local discriminator per site and per source-target configuration for improved performance. 

\par \textbf{Central:} A non-federated synthesis model was considered following data aggregation across sites \cite{pgan}. Loss function and architecture matched pFLSynth, albeit the mapper and PBs were excluded. The Central model serves as a privacy-violating benchmark for pFLSynth.

\par \textbf{FedGAN:}
A federated synthesis model was implemented with matching loss function and architecture to pFLSynth, but without the mapper and PBs \cite{FedGAN}. FedGAN aggregated the entire generator, so it serves as a non-personalized benchmark.

\par \textbf{FedMRI:}
A federated model proposed for maintaining site specificity in MRI reconstruction was considered \cite{fedmri}. FedMRI used a UNet backbone with a shared encoder and site-specific decoders \cite{fedmri}. For adversarial synthesis, a discriminator with matching loss function to pFLSynth was used. 

\par \textbf{FedMM:}
Federated implementation of a unified synthesis model (MM-GAN) was considered \cite{mmgan}. FedMM used a UNet generator and multiple input-output channels to cope with different source-target configurations \cite{mmgan}. The entire generator was aggregated. For fair comparison, curriculum learning was omitted to ensure standard sample selection.

\par \textbf{FedMed:}
A recent federated model for MRI synthesis was considered \cite{fedmedgan}. FedMed (abbreviated from FedMed-GAN) used a UNet generator that was entirely aggregated \cite{fedmedgan}. FedMed was originally proposed for unpaired synthesis. Only the forward mapping generator was retained for paired synthesis, and matching loss function to pFLSynth was used. For fair comparison, differential privacy procedures were omitted. 

\par \textbf{FedCycle:}
Originally proposed for low-dose CT denoising, FedCycle (abbreviated from FedCycleGAN) is a federated model based on a switchable backbone to reduce model complexity \cite{swfedcyclegan}. FedCycle was implemented with a UNet generator and adaptive normalization as in \cite{swfedcyclegan}. To adopt FedCycle for paired synthesis, only the forward mapping generator was retained, and matching loss function to pFLSynth was used. The switching mechanism was used to adapt the model to different source-target configurations as in pFLSynth.

\subsection{Architectural Details}
$M$ in pFLSynth had $L_M = 6$ layers. $S$ followed an encoder-bottleneck-decoder structure with $L_S=15$ stages. The encoder had 3 stages ($e1-e3$), each with a convolutional layer followed by a PB, and kernel sizes were 7, 3, 3 across stages. The bottleneck had 9 residual layers ($r1-r9$) with kernel size 3 \cite{resnet}, each residual layer followed by a PB. The decoder had 3 convolutional layers ($d1-d3$) of kernel sizes 3, 3, 7, with the first two layers followed by PBs. 
$D$ had 5 convolutional layers of kernel size 4. An FL setup with $K=4$ different datasets was considered, so a site index of $v^k \in \mathbb{Z}_2^{4}$ was used. Given multiple source-task configurations, uniform random sample selection was utilized for learning different tasks. The datasets examined included \Tone, \Ttwo, PD and FR contrasts, so a source-target configuration index of $u_c \in \mathbb{Z}_2^{8}$ was formed. The mapper received these indices and produced a latent vector $w^k_c \in \mathbb{R}^{512}$. The MLP in AdaCW had 2 layers with 64 hidden neurons. The splitting layer in the synthesizer ($L_U$) was selected as $r5$ based on validation performance. 


\subsection{Modeling Procedures}
For fair comparison, all competing models were trained using the same discriminator (Eq. \ref{eq:discrim}) and compound synthesis loss (Eq. \ref{eq:loss}). Hyperparameters including pixel-wise loss weight, number of communication rounds, number of epochs, and learning rate were selected via cross-validation. A common set of hyperparameters that yielded near-optimal results across models and datasets were selected. Training was performed via the Adam optimizer with batch size of 1 and $\beta_1 = 0.5$, $\beta_2 = 0.999$. For centralized models, training lasted 150 epochs. Training lasted $P$=150 rounds for federated models with $E$=1 local epochs each. Learning rate was set as $0.0002$ during the initial 75 epochs and linearly decayed to $0$ during the remaining epochs. The pixel-wise loss weight was set to $\lambda_{pix} = 100$. Models were implemented using PyTorch and Nvidia RTX 3090 GPUs. Performance was evaluated via PSNR, Structural Similarity Index (SSIM), and Fretchet Inception Distance (FID) \cite{fid_paper} metrics between synthetic and reference target-contrast images. Results were reported as mean and standard deviation across test subjects. Wilcoxon signed-rank tests were conducted to assess the significance of PSNR/SSIM differences among competing methods.

\begin{table*}[ht]
    \centering
    \footnotesize
    \resizebox{0.68\textwidth}{!}{
     \begin{tabular}{ l|c|c|c|c|c|c|c|c|c|c| }
        \cline{4-11}
        \multicolumn{3}{c|}{} & \multicolumn{2}{c|}{IXI} & \multicolumn{2}{c|}{BRATS} & \multicolumn{2}{c|}{MIDAS} & \multicolumn{2}{c|}{OASIS} \\
        \cline{4-11}
        \multicolumn{3}{c|}{} & \ToneTtwo & \TtwoPD & \ToneTtwo & \FLAIRTtwo & \ToneTtwo & \TtwoTone & \ToneTtwo & \TtwoFLAIR \\
        \cline{2-11}
        \multicolumn{1}{l|}{}& & P $\Uparrow$ & 25.2$\pm$1.3 & 26.8$\pm$1.8 & 24.5$\pm$1.0 & 21.8$\pm$1.5 & 22.7$\pm$1.0 & 20.8$\pm$1.0 & 24.3$\pm$1.2 & 21.9$\pm$1.4 \\
        \cline{3-11}
        \multicolumn{1}{l|}{}& Central & S $\Uparrow$ & 89.6$\pm$2.9 & 86.6$\pm$6.4 & 90.4$\pm$1.1 & 81.7$\pm$5.1 & 70.2$\pm$6.7 & 67.1$\pm$2.7 & 81.6$\pm$3.6 & 79.8$\pm$3.6 \\
        \cline{3-11}
        \multicolumn{1}{l|}{}& & F $\Downarrow$ & 29.1 & 59.1 & 45.3 & 85.1 & 48.0 & 11.5 & 39.1 & 26.4 \\
        \hline
        \multicolumn{1}{|l|}{\multirow{18}{*}{\rotatebox[origin=c]{90}{\textit{Federated models}}}} & & P $\Uparrow$ & \textbf{27.9$\pm$1.2} & \textbf{31.2$\pm$1.0} & \textbf{25.6$\pm$1.0} & \textbf{23.4$\pm$1.0} & \textbf{25.9$\pm$0.6} & 25.1$\pm$1.1 & \textbf{23.4$\pm$1.8} & \textbf{21.1$\pm$1.9} \\
        \cline{3-11}
        \multicolumn{1}{|l|}{}& pFLSynth & S $\Uparrow$ & \textbf{93.7$\pm$1.5} & \textbf{97.2$\pm$0.5} & \textbf{92.4$\pm$1.2} & \textbf{88.8$\pm$1.8} & \textbf{89.0$\pm$1.2} & 84.1$\pm$2.1 & \textbf{81.7$\pm$3.3} & \textbf{76.4$\pm$5.2} \\
        \cline{3-11}
        \multicolumn{1}{|l|}{}& & F $\Downarrow$ & \textbf{8.6} & \textbf{22.1} & \textbf{23.2} & \textbf{42.6} & \textbf{10.4} & \textbf{12.4} & 46.7 & \textbf{29.7} \\
        \cline{2-11}
        \multicolumn{1}{|l|}{}& & P $\Uparrow$ & 26.8$\pm$1.1 & 29.7$\pm$0.9 & 23.9$\pm$1.1 & 20.1$\pm$1.5 & 20.7$\pm$0.9 & 25.4$\pm$1.0 & 22.0$\pm$1.6 & 19.9$\pm$1.8 \\
        \cline{3-11}
        \multicolumn{1}{|l|}{}& FedGAN & S $\Uparrow$ & 92.2$\pm$1.7 & 96.2$\pm$0.6 & 89.9$\pm$1.5 & 81.5$\pm$2.7 & 76.4$\pm$3.2 & \textbf{85.5$\pm$2.1} & 78.2$\pm$3.3 & 75.2$\pm$5.5 \\
        \cline{3-11}
        \multicolumn{1}{|l|}{}& & F $\Downarrow$& 14.4 & 26.0 & 42.4 & 107.0 & 117.2 & 14.3 & 47.7 & 40.1 \\
        \cline{2-11}
        \multicolumn{1}{|l|}{}& & P $\Uparrow$ & 27.0$\pm$1.1 & 30.7$\pm$0.9 & 23.7$\pm$1.1 & 22.7$\pm$0.9 & 24.9$\pm$0.5 & 23.1$\pm$0.9 & 21.8$\pm$1.3 & 20.0$\pm$1.5 \\
        \cline{3-11}
        \multicolumn{1}{|l|}{}& FedMRI & S $\Uparrow$  & 92.7$\pm$1.5 & 96.6$\pm$0.5 & 89.8$\pm$1.7 & 87.8$\pm$1.6 & 86.7$\pm$1.2 & 79.0$\pm$2.2 & 77.7$\pm$2.7 & 72.3$\pm$5.0 \\
        \cline{3-11}
        \multicolumn{1}{|l|}{}& & F $\Downarrow$& 12.6 & 26.0 & 43.4 & 56.6 & 21.6 & 33.8 & \textbf{40.4} & 38.0 \\
        \cline{2-11}
        \multicolumn{1}{|l|}{}& & P $\Uparrow$ & 27.1$\pm$1.0 & 30.3$\pm$0.8 & 24.2$\pm$0.8 & 22.9$\pm$0.9 & 17.5$\pm$0.8 & 24.0$\pm$1.4 & 21.7$\pm$1.3 & 19.2$\pm$1.3 \\
        \cline{3-11}
        \multicolumn{1}{|l|}{}& FedMM & S $\Uparrow$ & 92.7$\pm$1.6 & 96.5$\pm$0.5 & 90.5$\pm$1.1 & 87.9$\pm$1.8 & 62.5$\pm$3.2 & 80.8$\pm$2.5 & 72.6$\pm$4.2 & 68.8$\pm$6.2 \\
        \cline{3-11}
        \multicolumn{1}{|l|}{}& & F $\Downarrow$& 13.4 & 24.8 & 36.4 & 46.6 & 169.6 & 26.9 & 45.4 & 39.7 \\
        \cline{2-11}
        \multicolumn{1}{|l|}{}& & P $\Uparrow$ & 26.2$\pm$0.9 & 28.7$\pm$0.8 & 22.8$\pm$1.1 & 20.1$\pm$1.0 & 25.5$\pm$0.7 & 24.0$\pm$1.4 & 21.4$\pm$1.2 & 19.6$\pm$1.8 \\
        \cline{3-11}
        \multicolumn{1}{|l|}{}& FedMed & S $\Uparrow$  & 91.5$\pm$1.6 & 95.1$\pm$0.7 & 88.2$\pm$1.6 & 82.3$\pm$2.0 & 88.9$\pm$1.2 & 80.8$\pm$2.5 & 75.7$\pm$3.3 & 71.2$\pm$5.3 \\
        \cline{3-11}
        \multicolumn{1}{|l|}{}& & F $\Downarrow$& 18.8 & 29.4 & 59.9 & 121.4 & 19.9 & 17.3 & 53.9 & 39.9 \\
        \cline{2-11}
        \multicolumn{1}{|l|}{}& & P $\Uparrow$ & 26.8$\pm$1.0 & 28.7$\pm$0.8 & 23.5$\pm$0.9 & 18.9$\pm$1.0 & 17.3$\pm$0.9 & \textbf{25.5$\pm$1.0} & 21.6$\pm$1.3 & 19.7$\pm$1.4 \\
        \cline{3-11}
        \multicolumn{1}{|l|}{}& FedCycle & S $\Uparrow$ & 92.5$\pm$1.5 & 95.6$\pm$0.7 & 89.3$\pm$1.0 & 80.1$\pm$2.7 & 62.0$\pm$3.2 & 84.0$\pm$2.2 & 65.7$\pm$5.1 & 64.0$\pm$8.4 \\
        \cline{3-11}
        \multicolumn{1}{|l|}{}& & F $\Downarrow$ & 16.5 & 31.0 & 56.1 & 139.1 & 176.8 & 14.0 & 49.8 & 47.9 \\
        \hline
    \end{tabular}
    }
    \caption{PSNR (P, dB), SSIM (S, \%), and FID (F) performance in a variable set of synthesis tasks. FR denotes FLAIR.}
    \label{tab:heterogeneous}
\end{table*}
\section{Results}

\subsection{Common Task Configuration}
\label{sec:first_setup}
FL experiments were conducted in a four-site setup based on IXI, BRATS, MIDAS, and OASIS datasets. We first demonstrated pFLSynth for decentralized modeling of a single synthesis task across sites (i.e., \ToneTtwon). This configuration reflects the influence of implicit heterogeneity in the data distribution due to cross-site differences in scanners and protocols used to collect the same MRI contrasts. Performance metrics for competing methods are listed in Table \ref{tab:unidirectional}. pFLSynth outperforms all federated baselines at each site (p$<$0.05), and offers on par performance with the central benchmark. On average across sites, pFLSynth achieves 0.8dB PSNR, 1.4\% SSIM, 5.8 FID improvement over competing FL baselines. Representative images are displayed in Fig. \ref{fig:exp_1}. pFLSynth reduces artifacts and noise, with the closest tissue depiction to the ground-truth target images, particularly near pathology.

\subsection{Variable Task Configuration}
\label{sec:second_setup}
We then demonstrated pFLSynth for simultaneous modeling of diverse tasks within and across sites (\ToneTtwo and \TtwoPD in IXI, \ToneTtwo and \FLAIRTtwo --FR denotes FLAIR-- in BRATS, \ToneTtwo and \TtwoTone in MIDAS, \ToneTtwo and \TtwoFLAIR in OASIS). This variable configuration creates explicit heterogeneity rendering suboptimal performance in a unified synthesis model, even for the central benchmark. Yet, we hypothesized that the personalized pFLSynth model should cope more reliably with heterogeneity. Performance metrics are listed in Table \ref{tab:heterogeneous}. pFLSynth outperforms all federated baselines at each site (p$<$0.05), except for MIDAS where FedCycle has higher PSNR/on par SSIM in \TtwoTonen, FedGAN has higher PSNR/SSIM in \TtwoTonen, FedMed has on par SSIM in \ToneTtwon, and for OASIS where FedMRI has lower FID in \ToneTtwon. On average across tasks and sites, pFLSynth achieves 2.0dB PSNR, 5.0\% SSIM, 24.9 FID improvement over federated baselines. pFLSynth also performs competitively against the central benchmark improving PSNR by 2.0dB, SSIM by 7.0\%, FID by 18.5. 
Representative images are shown in Fig. \ref{fig:exp_3}. Again, pFLSynth reduces artifacts and noise to accurately depict brain tissue. Altogether, the findings on common and variable configurations indicate the robustness of pFLSynth against implicit and explicit heterogeneity in MRI synthesis.

\begin{table}[t]
    \centering
    \footnotesize
    \resizebox{0.4\textwidth}{!}{ 
    \begin{tabular}{ |c|c|c|c|c|c| }
        \cline{3-6}
        \multicolumn{2}{c|}{} & IXI & BRATS & MIDAS & OASIS \\
        \cline{3-6}
        \multicolumn{2}{c|}{} & \ToneTtwo & \ToneTtwo & \ToneTtwo & \ToneTtwo \\
        \hline
        & P $\Uparrow$ & \textbf{28.6$\pm$1.3} & \textbf{26.3$\pm$1.0} & \textbf{28.2$\pm$0.5} & \textbf{25.1$\pm$0.6} \\
        \cline{2-6}
        pFLSynth & S $\Uparrow$ & \textbf{94.4$\pm$1.4} & \textbf{93.3$\pm$1.1} & \textbf{92.1$\pm$0.8} & \textbf{84.4$\pm$2.0} \\
        \cline{2-6}
        & F $\Downarrow$ & 7.7 & 23.9 & \textbf{9.2} & \textbf{30.2} \\
        \hline
        & P $\Uparrow$ & 28.2$\pm$1.3 & 25.9$\pm$0.9 & 28.0$\pm$0.5 & 24.6$\pm$0.7 \\
        \cline{2-6}
        w/o AdaIN & S $\Uparrow$ & 93.4$\pm$1.3 & 93.1$\pm$1.1 & 91.9$\pm$0.8 & 82.2$\pm$2.4 \\
        \cline{2-6}
        & F $\Downarrow$ & 7.9 & \textbf{23.8} & 9.8 & 32.7 \\
        \hline
        & P $\Uparrow$ & 28.2$\pm$1.2 & 26.2$\pm$0.8 & 28.1$\pm$0.5 & 25.0$\pm$0.7 \\
        \cline{2-6}
        w/o AdaCW & S $\Uparrow$ & 93.4$\pm$1.3 & 93.2$\pm$1.1 & 92.0$\pm$0.8 & 84.0$\pm$1.9 \\
        \cline{2-6}
        & F $\Downarrow$ & 8.6 & 24.6 & 9.5 & 32.0 \\
        \hline
        & P $\Uparrow$ & 28.0$\pm$1.2 & 26.0$\pm$0.9 & 28.2$\pm$0.5 & 24.9$\pm$0.7 \\
        \cline{2-6}
        w/o PNA & S $\Uparrow$ & 93.6$\pm$1.5 & 93.0$\pm$1.1 & 91.8$\pm$0.8 & 83.7$\pm$2.2 \\
        \cline{2-6}
        & F $\Downarrow$ & 10.2 & 24.6 & 9.3 & 35.6 \\
        \hline
        & P $\Uparrow$ & 28.4$\pm$1.2 & 9.9$\pm$0.5 & 28.1$\pm$0.5 & 24.8$\pm$0.6 \\
        \cline{2-6}
        w/o Mapper & S $\Uparrow$ & 94.2$\pm$1.3 & 72.6$\pm$1.8 & 92.0$\pm$0.9 & 83.1$\pm$2.0 \\
        \cline{2-6}
        & F $\Downarrow$ & \textbf{7.5} & 163.7 & 9.9 & 39.7 \\
        \hline
    \end{tabular}
    }
    \caption{Performance for pFLSynth and variants ablated of AdaIN, AdaCW, Partial Network Aggregation (PNA), and Mapper.}
    \label{tab:ablation1}
\end{table}

\begin{table*}[t]
    \centering
    \footnotesize
    \resizebox{0.65\textwidth}{!}{
     \begin{tabular}{ |c|c|c|c|c|c|c|c|c|c| }
        \cline{3-10}
        \multicolumn{2}{c|}{} & \multicolumn{2}{c|}{IXI} & \multicolumn{2}{c|}{BRATS} & \multicolumn{2}{c|}{MIDAS} & \multicolumn{2}{c|}{OASIS} \\
        \cline{3-10}
        \multicolumn{2}{c|}{} & \ToneTtwo & \TtwoPD & \ToneTtwo & \FLAIRTtwo & \ToneTtwo & \TtwoTone & \ToneTtwo & \TtwoFLAIR \\
        \hline
        & P $\Uparrow$ & \textbf{27.9$\pm$1.2} & \textbf{31.2$\pm$1.0} & \textbf{25.6$\pm$1.0} & \textbf{23.4$\pm$1.0} & \textbf{26.0$\pm$0.6} & 25.1$\pm$1.1 & \textbf{23.4$\pm$1.8} & \textbf{21.1$\pm$1.9} \\
        \cline{2-10}
        pFLSynth & S $\Uparrow$ & \textbf{93.7$\pm$1.5} & \textbf{97.2$\pm$0.5} & \textbf{92.4$\pm$1.2} & \textbf{88.8$\pm$1.8} & 89.0$\pm$1.2 & 84.1$\pm$2.1 & \textbf{81.7$\pm$3.3} & \textbf{76.4$\pm$5.2} \\
        \cline{2-10}
        & F $\Downarrow$ & \textbf{8.6} & \textbf{22.1} & \textbf{23.2} & \textbf{42.6} & \textbf{10.4} & \textbf{12.4} & \textbf{46.7} & \textbf{29.7} \\
        \hline
        & P $\Uparrow$ & 27.6$\pm$1.2 & 30.9$\pm$1.0 & 25.2$\pm$1.0 & 18.5$\pm$1.0 & 25.6$\pm$0.6 & 23.8$\pm$1.0 & 23.2$\pm$1.7 & 19.7$\pm$1.9 \\
        \cline{2-10}
        w/o $v^k$ & S $\Uparrow$ & 93.4$\pm$1.4 & 97.0$\pm$0.4 & 91.5$\pm$1.1 & 80.5$\pm$2.0 & \textbf{89.2$\pm$1.2} & 82.2$\pm$1.2 & 74.9$\pm$4.5 & 71.1$\pm$6.3 \\
        \cline{2-10}
        & F $\Downarrow$ & 13.7 & 22.5 & 36.2 & 147.9 & 17.8 & 25.0 & 49.4 & 39.7 \\
        \hline
        & P $\Uparrow$ & 27.4$\pm$1.2 & 30.3$\pm$0.9 & 24.6$\pm$1.1 & 7.1$\pm$8.7 & 25.8$\pm$0.6 & \textbf{25.2$\pm$1.0} & 22.2$\pm$1.6 & 20.5$\pm$2.0 \\
        \cline{2-10}
        w/o $u_c$ & S $\Uparrow$ & 93.6$\pm$1.4 & 97.0$\pm$0.4 & 91.2$\pm$1.2 & 29.7$\pm$2.3 & 88.9$\pm$1.1 & \textbf{84.7$\pm$2.1} & 80.5$\pm$3.1 & 74.5$\pm$5.4 \\
        \cline{2-10}
        & F $\Downarrow$ & 9.8 & 22.7 & 23.5 & 162.8 & 10.6 & 13.0 & 47.3 & 31.1 \\
        \hline
    \end{tabular}
    }
    \caption{Performance for pFLSynth and variant models ablated of site index ($v^k$) and source-target configuration index ($u_c$).}
    \label{tab:ablation2}
\end{table*}

\subsection{Ablation Studies}
Ablation studies were conducted to assess the contributions of major design elements in pFLSynth. First, pFLSynth was compared against variants where AdaIN layers were ablated, AdaCW layers were ablated, full network aggregation was used, and the mapper was ablated (i.e., site and source-target configuration indices directly input to PBs). Performance metrics for the common task configuration are listed in Table \ref{tab:ablation1}. pFLSynth outperforms all ablated variants at each site, except for the mapper-ablated variant that yields slightly lower FID in IXI, and the AdaIN-ablated variant that yields slightly lower FID in BRATS. These results demonstrate the importance of PBs, partial network aggregation, and the mapper to synthesis performance. 
We then compared pFLSynth against variants where site index was removed, and source-target configuration index was removed. Performance in the variable task configuration is summarized in Table \ref{tab:ablation2}. pFLSynth outperforms ablated variants across sites and tasks, except for occasionally albeit slightly higher PSNR and SSIM with variants in MIDAS. These results demonstrate the importance of using personalized latents per site and source-target configuration.

Next, we examined the effect of delayed insertion of a specific site or translation task to the FL setup. To do this, spare digits were reserved in the site index and source-target configuration index to code late joiners included halfway during the training. For delayed site analysis, a single site was held out in the common task configuration and pFLSynth was compared to a variant with ablated site index. For delayed task analysis, a single task was held out in the variable task configuration and pFLSynth was compared to a variant with ablated source-target configuration index. Performance metrics for the held-out site and task are listed in Table \ref{tab:late_join} (there were unsubstantial differences for non-held-out sites/tasks). Models with delayed site or task perform competitively with the original model including all sites and tasks, and they outperform variants with ablated site or source-target configuration index. These results suggest that pFLSynth shows a degree of inherent reliability against delayed insertion. 

\begin{figure}[t]
\centering 
\includegraphics[width=\columnwidth]{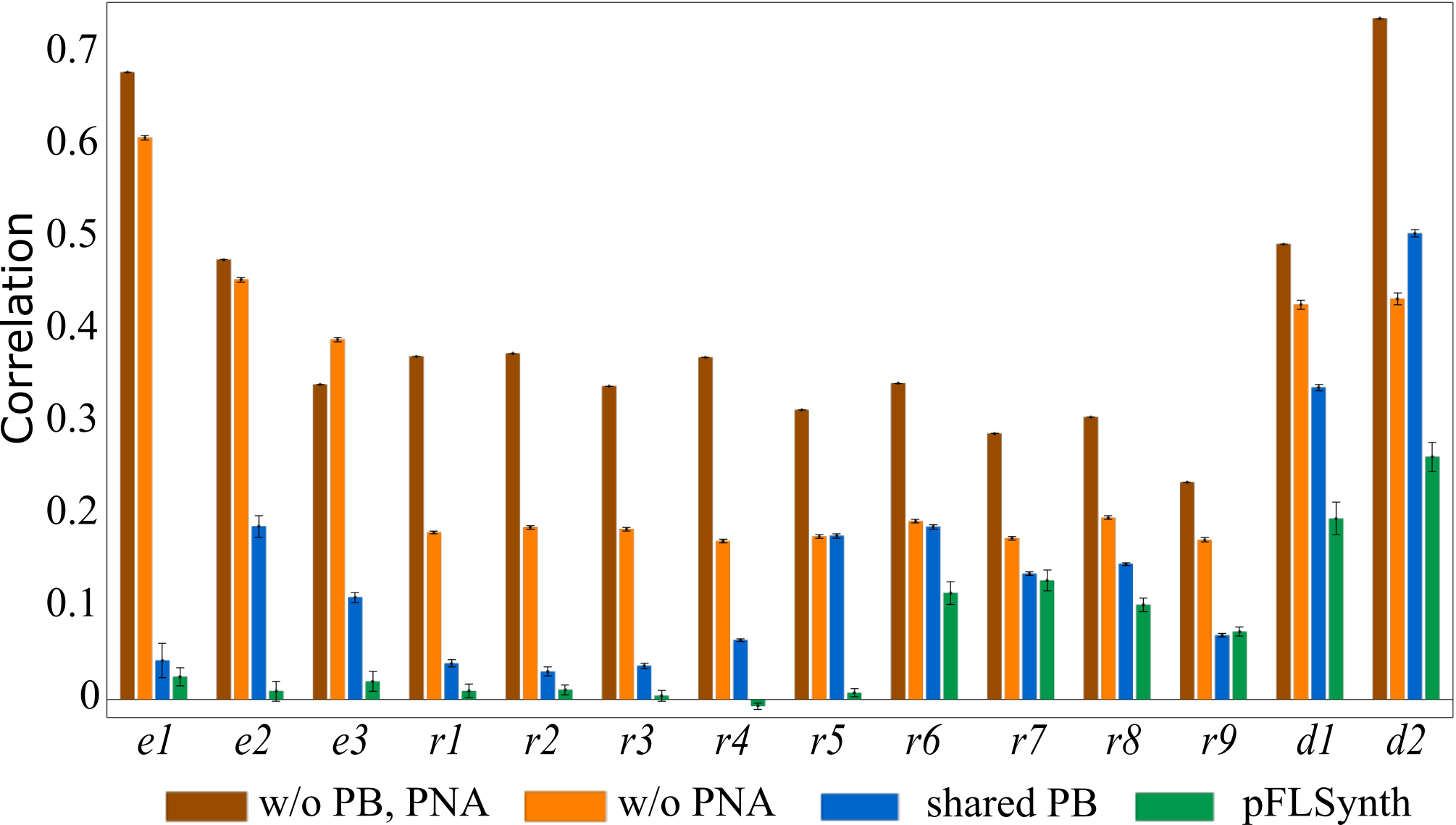}
\caption{Similarity of activation maps across sites in different stages of the synthesizer. A set of training images were projected through the local synthesizers at each site, and the resultant activation maps were compared across sites. Results are shown as mean$\pm$std of Spearman's correlation coefficient across sites.}
\label{fig:similarity}
\end{figure}

Lastly, we assessed the influence of PBs and PNA on possible information leakage among FL sites. Recent studies posit layer-wise measures to assess leakage in network models \cite{DBLP:journals/corr/Shwartz-ZivT17}. Thus, we measured the similarity of activation maps in local synthesizer layers to assess the potential for leakage \cite{DBLP:journals/corr/abs-2010-08762}. A random set of 50 training source images were selected from each site and projected separately through all local synthesizers. Similarity for a given source image was taken as Spearman's correlation coefficient between the activation maps it elicits in separate sites. Fig. \ref{fig:similarity} displays similarity for pFLSynth, variants with ablated PBs and/or PNA, and a variant that shared PBs across sites. Similarity of activation maps is lowered by inclusion of both PBs and PNA. Unshared PBs in pFLSynth further reduce similarity in activation maps, suggesting enhanced reliability against leakage. 

\begin{table}[]
\centering
\begin{tabular}{lcc|c|cc|c|}
\cline{4-4} \cline{7-7}
                                              &                                                                                                             &                & MIDAS               &                                                                                                              &                & IXI                 \\ \cline{4-4} \cline{7-7} 
                                              &                                                                                                             &                & \ToneTtwo &                                                                                                              &                & \TtwoPD \\ \cline{2-7} 
\multicolumn{1}{l|}{}                         & \multicolumn{1}{c|}{\multirow{3}{*}{\begin{tabular}[c]{@{}c@{}}pFLSynth\\ (original)\end{tabular}}}         & P $\Uparrow$   & 25.1$\pm$0.6           & \multicolumn{1}{c|}{\multirow{3}{*}{\begin{tabular}[c]{@{}c@{}}pFLSynth\\ (original)\end{tabular}}}          & P $\Uparrow$   & 31.2$\pm$1.0           \\ \cline{3-4} \cline{6-7} 
\multicolumn{1}{l|}{}                         & \multicolumn{1}{c|}{}                                                                                       & S $\Uparrow$   & 84.4$\pm$2.0           & \multicolumn{1}{c|}{}                                                                                        & S $\Uparrow$   & 97.2$\pm$0.5           \\ \cline{3-4} \cline{6-7} 
\multicolumn{1}{l|}{}                         & \multicolumn{1}{c|}{}                                                                                       & F $\Downarrow$ & 30.2                & \multicolumn{1}{c|}{}                                                                                        & F $\Downarrow$ & 22.1                \\ \cline{1-7} 
\multicolumn{1}{|l|}{\multirow{6}{*}{\rotatebox[origin=c]{90}{Delayed insertion}}} & \multicolumn{1}{c|}{\multirow{3}{*}{\begin{tabular}[c]{@{}c@{}}pFLSynth \\ (delayed \\ site)\end{tabular}}} & P $\Uparrow$   & 24.8$\pm$0.8           & \multicolumn{1}{c|}{\multirow{3}{*}{\begin{tabular}[c]{@{}c@{}}pFLSynth \\ (delayed \\ task)\end{tabular}}}  & P $\Uparrow$   & 30.9$\pm$1.0           \\ \cline{3-4} \cline{6-7} 
\multicolumn{1}{|l|}{}                         & \multicolumn{1}{c|}{}                                                                                       & S $\Uparrow$   & 82.9$\pm$2.2           & \multicolumn{1}{c|}{}                                                                                        & S $\Uparrow$   & 97.0$\pm$0.5           \\ \cline{3-4} \cline{6-7} 
\multicolumn{1}{|l|}{}                         & \multicolumn{1}{c|}{}                                                                                       & F $\Downarrow$ & 37.6                & \multicolumn{1}{c|}{}                                                                                        & F $\Downarrow$ & 23.9                \\ \cline{2-7} 
\multicolumn{1}{|l|}{}                         & \multicolumn{1}{c|}{\multirow{3}{*}{\begin{tabular}[c]{@{}c@{}}w/o $v_k$\\ (delayed \\ site)\end{tabular}}}    & P $\Uparrow$   & 23.0$\pm$1.1           & \multicolumn{1}{c|}{\multirow{3}{*}{\begin{tabular}[c]{@{}c@{}}w/o $u_c$ \\ (delayed\\  task)\end{tabular}}} & P $\Uparrow$   & 30.5$\pm$0.9           \\ \cline{3-4} \cline{6-7} 
\multicolumn{1}{|l|}{}                         & \multicolumn{1}{c|}{}                                                                                       & S $\Uparrow$   & 68.8$\pm$3.3           & \multicolumn{1}{c|}{}                                                                                        & S $\Uparrow$   & 96.6$\pm$0.5           \\ \cline{3-4} \cline{6-7} 
\multicolumn{1}{|l|}{}                         & \multicolumn{1}{c|}{}                                                                                       & F $\Downarrow$ & 44.6                & \multicolumn{1}{c|}{}                                                                                        & F $\Downarrow$ & 24.7                \\ \cline{1-7} 
\end{tabular}
\caption{Effects of delayed insertion. Results shown for the original pFLSynth with all sites/tasks included, pFLSynth with delayed site/task insertion, and variant models ablated of site or task index.}
\label{tab:late_join}
\end{table}

\subsection{Complexity and Communication Efficiency}
A practical concern for FL models is efficiency in decentralized settings. For each model, Table \ref{tab:complexity_communication_load} lists number of total versus communicated parameters. While pFLSynth has moderately higher complexity, it has comparable training and inference times to other models (unreported). Importantly, as pFLSynth does not communicate upstream generator layers, it performs competitively in communication efficiency, showcasing an added benefit of PNA.


\section{Discussion}
Federated MRI synthesis has to operate under distributional heterogeneity in multi-site imaging data \cite{li2021fedbn}. A recent study has considered FedAvg optimization of cycle-consistent models for MRI synthesis \cite{fedmedgan}. However, no prior study has proposed a dedicated mechanism to address data heterogeneity beyond the level inherently offered by FedAvg. To our knowledge, pFLSynth is the first FL method to personalize a global synthesis model to each individual site and translation task. Experiments on multi-site MRI data demonstrate that pFLSynth offers on par performance to a central benchmark while outperforming federated baselines. Our results suggest that pFLSynth can improve generalizability and flexibility in multi-site collaborations by permitting training on imaging data from diverse sources and protocols.

\par
FL avoids transfer of imaging data to mitigate patient privacy risks. Yet, inference attacks might leak information about training data from model parameters \cite{Kaissis2020}. Here, we considered an FL setup where only generators were shared across sites. Yet, the discriminators were never communicated since our experiments in the initial phases of the study indicated that sharing discriminators did not elicit any performance benefit. This setup is reported to be relatively resilient against inference attacks \cite{Han2020}. Nevertheless, potential risks can be further minimized by adopting differentially private training \cite{fedmedgan,ziller2021}, or by extending the size and diversity of the training datasets to implicitly improve privacy \cite{FengICCV2021}. Future studies are warranted to systematically examine the privacy properties of FL-based methods in multi-contrast MRI synthesis. 

 \begin{table}[t]
     \centering
     \footnotesize
     \resizebox{\columnwidth}{!}{
     \begin{tabular}{ |c|c|c|c|c|c|c|c| }
         \cline{2-8}
         \multicolumn{1}{c|}{} & Central & pFLSynth & FedGAN & FedMRI & FedMM & FedMed & FedCycle \\
         \hline
         Comp.& 14.27 & 18.75 & 14.27 & 10.51 & 10.52 & 10.51 & 12.75 \\
         \hline
         Comm.& - & 6.52 & 11.51 & 4.71 & 7.77 & 7.77 & 9.99 \\
         \hline
     \end{tabular}
     }
     \caption{Local model complexity (millions of parameters) and communication load (millions of parameters) for competing methods.}
     \label{tab:complexity_communication_load}
 \end{table}

Here, we primarily considered an FL setup where all sites contributed to the entire training process. When late-joining sites were present, successful learning was achieved given a reasonable number of communication rounds after inclusion. An alternative is to use local mappers at each site to learn site-specific latents without needing a site index. Certain scenarios might involve inference at a site that was completely excluded from training. In those cases, an average model across training sites might yield suboptimal performance at a held-out site. To improve performance, transfer-learning procedures can be adopted to fine-tune the trained model on a compact dataset from the held-out site \cite{pgan}. 

The proposed method can be developed along several technical lines. We performed training on paired datasets with registered source and target images from a matching set of subjects. Utilization of unpaired data can facilitate compilation of broader datasets for training substantially complex models. In those cases, cycle-consistent \cite{woltering2017,ge2019} or semi-supervised training \cite{jin2018} strategies can be adopted. Unpaired learning might further be enhanced via dedicated spatial alignment modules \cite{fedmedatl}. We demonstrated pFLSynth for one-to-one tasks with a single source and a single target contrast. When information to synthesize the target modality is not sufficiently evident in a single source modality, pFLSynth can be generalized to perform many-to-one mapping \cite{mustgan,collagan,mmgan}. Recent studies have reported benefits for training centralized models based on transformers instead of CNNs \cite{resvit,SLATER}. Relatively light transformer modules might be preferable to maintain reasonable computation and communication costs.

\section{Conclusion}
We introduced a novel personalized FL method for multi-contrast MRI synthesis based on an adversarial model equipped with personalization blocks and partial network aggregation. Benefits over prior federated methods were demonstrated for brain image synthesis considering different public datasets and source-target configurations. Improved generalization against implicit and explicit domain shifts renders pFLSynth a promising candidate for multi-site collaborations. pFLSynth might also be adopted for other image translation tasks involving CT or PET, and other dense-prediction tasks such as reconstruction or super-resolution.

\bibliographystyle{IEEEtran}
\bibliography{IEEEabrv,references}
\end{document}